\documentclass[onecolumn,a4paper,12pt,oneside]{article} 
\usepackage[top=2.5cm,left=2.5cm,right=2.5cm,bottom=3cm]{geometry}    
\usepackage[nostamp]{draftwatermark} % watermark programmed but not printed
\SetWatermarkAngle{45}
\SetWatermarkFontSize{1.3cm}
\SetWatermarkScale{2.5}
\SetWatermarkLightness{0.85}     
\usepackage[parfill]{parskip}    % Activate to begin paragraphs with an empty line rather than an indent
\usepackage{graphicx}
\usepackage{amssymb}
\usepackage{amsmath}
\usepackage{amsfonts}
\usepackage{amsthm}
\usepackage{epstopdf}
\usepackage[usenames,dvipsnames,svgnames,table]{xcolor}
\usepackage{tikz}
\usepackage{array}
\usepackage[T1]{fontenc}
\usepackage[utf8]{inputenc}
\usepackage{booktabs} % for much better looking tables
\usepackage{array} % for better arrays (eg matrices) in maths
\usepackage{paralist} % very flexible & customisable lists (eg. enumerate/itemize, etc.)
\usepackage{listings} 
\usepackage{verbatim} % adds environment for commenting out blocks of text & for better verbatim
\usepackage{subfig} % make it possible to include more than one captioned figure/table in a single float
\usepackage[colorlinks=true]{hyperref}
\hypersetup{colorlinks=true, linktocpage=true, linkcolor=Blue, citecolor=Green,
hypertexnames=true, urlcolor=Blue,pdftex}
\usepackage[english]{babel}
\usepackage{wrapfig}
\usepackage{multirow}
\usepackage{quoting}
\usepackage{rotating}
\quotingsetup{font=normalsize}
\theoremstyle{plain}

\usepackage{bm}
\usepackage{abstract}
\usepackage{cite}
\usepackage{float}
\usepackage{textcomp} 	
\usepackage{tabularx} % Better tables
\usepackage{caption}
\usepackage{authblk} % for authors, affiliations etc.
\usepackage{eso-pic} % for background figures
\usepackage{enumitem} % betters lists
\usepackage{lineno} % for numbering lines
\providecommand{\keywords}[1]{\textbf{{Key words: }} #1} % command for keywords

\usepackage{graphicx}
\usepackage{multirow}
\usepackage{amsmath,amssymb,amsfonts}
\usepackage{mathrsfs}
\usepackage[title]{appendix}
\usepackage{xcolor}
\usepackage{textcomp}
\usepackage{manyfoot}
\usepackage{booktabs}
\usepackage{algorithm}
\usepackage{algorithmicx}
\usepackage{algpseudocode}
\usepackage{listings}
\usepackage{paralist}

\newcommand{\be}{\begin{equation}}
\newcommand{\ee}{\end{equation}}
\newcommand{\bs}{\begin{split}}
\newcommand{\es}{\end{split}}

\renewcommand{\Phi}{\varPhi}

\newcommand{\Sy}{\mathbb{S}}

\renewcommand{\Theta}{\varTheta}
\renewcommand{\Psi}{\varPsi}
\renewcommand{\Sigma}{\varSigma}
\newcommand{\A}{\mathbb{A}}

\newcommand{\B}{\mathbb{B}}

\newcommand{\D}{\mathbb{D}}

\newcommand{\Q}{\mathbb{Q}}

\renewcommand{\Delta}{\varDelta}
\renewcommand{\phi}{\varphi}
\renewcommand{\psi}{\varPsi}

\newcommand{\nud}{\nu_{12}}

\title{\LARGE{\bf Optimizing the auxetic behavior of anisotropic laminates}}

\author{Paolo Vannucci\smallskip\\
\begin{small}{ LMV - Laboratoire de Mathématiques de Versailles, UMR8100 \\
 Université de Versailles et Saint Quentin - 45, Avenue des Etats-Unis, 78035 - France\\
           \href{mailto:paolo.vannucci@uvsq.fr}{paolo.vannucci@uvsq.fr}} \end{small} \bigskip\\
             - PREPRINT -  \bigskip\bigskip}

\begin{document}

\maketitle

\begin{abstract}
{Anisotropic laminates with  a negative Poisson's ratio for at least some directions are called auxetic. In this paper, we consider the conditions for optimizing the auxeticity of an orthotropic laminate, namely: for a laminate composed by a given material, (i) how to obtain the lowest, i.e. the highest negative, Poisson's ratio  and (ii) how to maximize the auxetic zone, i.e. the set of directions where the Poisson's ratio is negative. It is shown that in both the cases the optimal solution is found on the boundary of the feasible domain and in particular that it can be obtained using angle-ply sequences of identical layers.  The polar method with dimensionless moduli is employed for representing the anisotropic behavior of the laminate, which allows, on the one hand, to reduce the dimensionality of the problem and, on the other hand, to have an effective mathematical representation of anisotropy by dimensionless invariants.}
\end{abstract}

\keywords{Poisson's ratio, auxeticity, anisotropy, composite laminates, polar formalism, angle-ply laminates, optimization}

\section{Introduction}
Auxeticity is the property of having a negative Poisson's ratio, \cite{cho2019}. It is well known that an auxetic behavior is theoretically possible for isotropic materials, see e.g. \cite {Love, Sokolnikoff, Gurtin}, and that it can be obtained using materials having some special kind of microstructure (literature is very wide on this topic, some rather complete reviews on the matter can be found in \cite{Prawoto12} or \cite{Shukla22}). In the case of anisotropic materials, the Poisson's coefficients are anisotropic properties too and by consequence it is physically possible that they take negative values at some directions. In other words, auxeticity of anisotropic materials is an anisotropic property. When anisotropic layers are used in the fabrication of laminates, the mechanical properties of theses ones can be tailored by an appropriate choice of the orientation angles of the plies, see e.g. \cite{Jones,gay14,vasiliev}. Hence, also auxeticity can be tailored, at least for some constituent materials. The possibility of obtaining auxetic orthotropic laminates has been treated in  \cite{vannucci24b}, where it has been shown which are the conditions on the mechanical properties of the constituent layer that are needed to fabricate, by an appropriate tailoring of the orientations,  a laminate that has, for its in-plane behavior, a negative Poisson's ratio at some directions. 

In this paper, we complete the study considering two optimization problems concerning auxetic orthotropic laminates: for a given constituent layer, what are the conditions to obtain (i) the lowest, negative, in-plane Poisson's ratio $\nud$ or (ii) the largest auxetic zone, i.e. the set of directions where $\nud<0$? The first problem has already been considered by Miki and Murotsu, in a seminal paper on auxetic laminates, \cite{Miki89}, and later on by  Zhang et al., \cite{Zhang98}. Miki and Murotsu used a numerical optimization technique  and argued that the maximization of $\nud$, that in their paper is indicated by $\nu_{21}$, can be obtained using angle-ply laminates, while  the minimum is obtained still using two different orientations, but with unbalanced laminates, that  hence are not orthotropic. Zhang et al. considered exclusively in-plane orthotropic laminates, assumed, without proving, that in this case the minimum of $\nud$ can be attained using angle-ply laminates and gave an approximated analytical expression for the gap angle between the two sets of layers for a minimization problem not clearly defined. The second problem is apparently still untreated in the literature. 

In this paper we use the polar formalism, introduced in 1979 by G. Verchery, \cite{verchery79}, particularly interesting as it makes use of invariants and of angles to represent the elastic properties. This method is very effective also in exploring unconventional mechanical situations, like the case of auxetic materials or also other more strange situations, see e.g. \cite{MMAS12a}. In particular, like in \cite{vannucci24b}, we introduce dimensionless polar parameters, which allows to reduce the mathematical dimensionality  of the problem. We   show that the minimum $\nud$ and the widest angular range where $\nud<0$ can be obtained using angle-ply laminates  and we give also an analytical expression for the optimal solution of the second problem.

\section{Basic relations}
\subsection{Polar expression of the Poisson's ratio}
We consider an anisotropic ply in a  plane-stress state, we fix a frame $\{x_1,x_2\}$ and denote by $\Sy$ and $\Q=\Sy^{-1}$ respectively the compliance and the reduced stiffness  tensors. Then, \cite{Lekhnitskii,TsaiHahn,Jones,Ting}, 
\be
\label{eq:nud}
\nud(\theta):=-\frac{\Sy_{12}(\theta)}{\Sy_{11}(\theta)}
\ee
is the in-plane Poisson's ratio for the direction inclined of $\theta$ on the $x_1-$axis. 
Because $\Sy_{11}(\theta)>0\ \forall\theta$, 
\be
\nud(\theta)<0\iff\Sy_{12}(\theta)>0.
\ee
In the polar formalism, \cite{vannucci05,vannucci_libro}
\be
\label{eq:auxeticity0}
\nud(\theta)=\frac{t_0-2t_1+ r_0\cos4(\phi_0-\theta)}{t_0+2t_1+ r_0\cos4(\phi_0-\theta)+4r_1\cos2(\phi_1-\theta)},
\ee
with $t_0, t_1,r_0,r_1$ non-negative tensor invariants  of $\Sy$ while $\phi_0,\phi_1$ are polar angles whose difference is the fifth invariant of $\Sy$. Because we investigate the auxeticity of a laminate, whose elastic behavior is obtained homogenizing the reduced stiffnesses of the layers, it is worth to express $\nud(\theta)$ as function of the polar invariants of $\Q$. To this end, we  express  the polar parameters of  $\Sy$ by those of $\Q$, \cite{vannucci_libro}:
\be
\begin{array}{l}
t_0=2\dfrac{T_0T_1-R_1^2}{\Delta},\medskip\\
t_1=\dfrac{T_0^2-R_0^2}{2\Delta},\medskip\\
r_0 \mathrm{e}^{4\mathrm{i}\phi_0}=\dfrac{2}{\Delta}(R_1^2\mathrm{e}^{4\mathrm{i}\Phi_1}-T_1R_0\mathrm{e}^{4\mathrm{i}\Phi_0}),\medskip\\
r_1 \mathrm{e}^{2\mathrm{i}\phi_1}=\dfrac{R_1\mathrm{e}^{2\mathrm{i}\Phi_1}}{\Delta}\left[R_0\mathrm{e}^{4\mathrm{i}\Phi}-T_0\right].
\end{array}
\ee
Here, $T_0,T_1,R_0,R_1$ are the (non-negative) polar invariant moduli of $\Q$ and $\Phi_0,\Phi_1$ the two polar angles of $\Q$, whose difference is 
\be
\Phi:=\Phi_0-\Phi_1,
\ee
which  is the fifth independent invariant of $\Q$. $\Delta$ is the  quantity 
\be
\Delta=4T_1(T_0^2-R_0^2)-8R_1^2[T_0-R_0\cos4\Phi],
\ee
which actually coincides with the determinant of $\Q$, and as such is itself a positive invariant.
Some simple passages give
\be
\label{eq:nud1}
\hspace{-10mm}\nud(\theta)=\frac{2(T_0T_1{-}R_1^2){-}T_0^2{+}R_0^2{+}2[R_1^2\cos4(\Phi_1{-}\theta){-}T_1R_0\cos4(\Phi_0{-}\theta)]}{2(T_0T_1{-}R_1^2){+}T_0^2{-}R_0^2{+}2[R_1^2\cos4(\Phi_1{-}\theta){-}T_1R_0\cos4(\Phi_0{-}\theta)]{+}4R_1(R_0\mathrm{e}^{4\mathrm{i}\Phi}{-}T_0)\cos2(\Phi_1{-}\theta)}.
\ee

We assume that the ply is a unidirectional (UD) layer, so an ordinary orthotropic planar material. This means that  $R_0\neq0,R_1\neq0$, \cite{vannucci02joe,vincenti01}, and that, \cite{vannucci05}, 
\be
\Phi=K\frac{\pi}{4},\ \ K\in\{0,1\}.
\ee
In the polar formalism, any rotation  of the frame corresponds to subtract the rotation angle from the two polar angles. If the rotation amounts to $\Phi_1$ or, equivalently, if the reference frame is chosen in such a way that $\Phi_1=0$, which corresponds, for an UD ply, to put the $x_1-$axis aligned with the fibres, eq. (\ref{eq:nud1}) becomes
\be
\label{eq:nud2}
\hspace{-4mm}\nud(\theta)=\frac{2(T_0T_1{-}R_1^2){-}T_0^2{+}R_0^2{+}2[R_1^2{-}(-1)^KT_1R_0]\cos4\theta}{2(T_0T_1{-}R_1^2){+}T_0^2{-}R_0^2{+}2[R_1^2{-}(-1)^KT_1R_0]\cos4\theta{+}4R_1[(-1)^KR_0{-}T_0]\cos2\theta}.
\ee

We focus on the laminate's extension response, described by tensor\cite{Jones,vannucci_libro}
\begin{equation}
\label{eq:Atens}
\A=\frac{1}{h}\sum_{j=1}^n({z_j}-{z_{j-1}})\Q_(\delta_j),
\end{equation}
of a laminate composed of identical UD layers having a reduced stiffness tensor $\Q$. In the above equation, 
 $\delta_j$ is the orientation of the $j-$th layer among the $n$ composing the laminate while $h$ is the plate's thickness. 
 
Moreover, we assume that the laminate is extension-bending uncoupled. This assumption is important because it is practically impossible to express analytically the Poisson's ratio of a coupled laminate, as in such a case the compliances, in extension and in bending, depend in a very complicate manner upon $\A,\B$ and $\D$, respectively the stiffness tensors in extension, coupling and bending\cite{vannucci01joe,vannucci23a,vannucci23b}. Here, it is interesting to recall that uncoupling, i.e. $\B=\mathbb{O}$,  contrarily to what commonly written, is not necessarily get using symmetric stacks. In \cite{vannucci01ijss,vannucci01cst}, it has been shown that asymmetric uncoupled laminates are much more numerous than the symmetric ones. Finally,  uncoupling can be obtained rather easily and  it is not limitative to assume it.

The expression of $\nud(\theta)$ for an orthotropic tensor $\A$  is the same of that for $\Q$, provided that the polar parameters of $\A$ (denoted in the following  by a superscript $A$) replace those of the layer. For a laminate composed of identical plies,
 the polar parameters of  tensor $\A$  are related to those of the constituent layer by, \cite{vannucci_libro},
\begin{equation}
\label{eq:compApolareq}
\begin{array}{l}
T_0^A=T_0,\medskip\\
T_1^A=T_1,\medskip\\
R_0^A\mathrm{e}^{4\mathrm{i}\varPhi_0^A}={R_0\mathrm{e}^{4\mathrm{i}\varPhi_0}}{\left(\xi_1+\mathrm{i}\xi_2\right)},\medskip\\
R_1^A\mathrm{e}^{2\mathrm{i}\varPhi_1^A}={R_1\mathrm{e}^{2\mathrm{i}\varPhi_1}}{\left(\xi_3+\mathrm{i}\xi_4\right)}.
\end{array}
\end{equation}
The quantities $\xi_i,i=1,...,4$ are the {\it lamination parameters}\cite{tsai1968} for $\A$:
\begin{equation}
\label{eq:laminationparameters}
\xi_1+\mathrm{i}\xi_2=\frac{1}{n}\sum_{j=1}^n\mathrm{e}^{4\mathrm{i}\delta_j},\ \ \xi_3+\mathrm{i}\xi_4=\frac{1}{n}\sum_{j=1}^n\mathrm{e}^{2\mathrm{i}\delta_j},
\end{equation}
We focus on laminates with $\A$ orthotropic; then,  choosing $\Phi^A_1=0$ to fix the reference frame for the laminate, we get easily
\be
%\begin{array}{c}
\xi_2=\xi_4=0,\  \ (-1)^{K^A}R_0^A=(-1)^KR_0\xi_1,\ \ R_1^A=R_1\xi_3.
%\end{array}
\ee
So, finally
\be
\label{eq:nud3}
\hspace{-8mm}\nud^A(\theta)=\frac{2(T_0T_1{-}R_1^2\xi_3^2){-}T_0^2{+}R_0^2\xi_1^2{+}2[R_1^2\xi_3^2{-}(-1)^KT_1R_0 \xi_1]\cos4\theta}{2(T_0T_1{-}R_1^2\xi_3^2){+}T_0^2{-}R_0^2\xi_1^2{+}2[R_1^2\xi_3^2{-}(-1)^KT_1R_0\xi_1]\cos4\theta{+}4R_1\xi_3[(-1)^KR_0\xi_1{-}T_0]\cos2\theta}.
\ee

In this equation,  the independent variable $\theta$ determines the {\it direction}, the two lamination parameters $\xi_1$ and $\xi_3$ account for the {\it geometry}  of the stack (namely, for the sequence of the  angles $\delta_j$), while the polar parameters $T_0,T_1,(-1)^KR_0$ and $R_1$ of the layer represent the {\it material} part of $\nud^A(\theta)$.

We introduce now three dimensionless parameters for the material part, \cite{vannucci13}:
\be
\label{eq:adimpar}
\tau_0=\frac{T_0}{R_1},\ \ \tau_1=\frac{T_1}{R_1},\ \ \rho=\frac{R_0}{R_1}.
\ee

This allows the reduction of  the number of independent material parameters and  an easier study of the relation between geometry and material. It is worth noting that these ratios can be introduced because, for a UD ply, $R_1\neq0$. Then, Eq. (\ref{eq:nud3}) becomes
\be
\label{eq:nud4}
\hspace{-2mm}\nud^A(\theta)=\frac{2(\tau_0\tau_1{-}\xi_3^2){-}\tau_0^2{+}\rho^2\xi_1^2{+}2[\xi_3^2{-}(-1)^K\tau_1\rho \xi_1]\cos4\theta}{2(\tau_0\tau_1{-}\xi_3^2){+}\tau_0^2{-}\rho^2\xi_1^2{+}2[\xi_3^2{-}(-1)^K\tau_1\rho\xi_1]\cos4\theta{+}4\xi_3[(-1)^K\rho\xi_1{-}\tau_0]\cos2\theta}.
\ee

\subsection{The lamination domain}
\begin{figure}
\centering
\includegraphics[width=.5\columnwidth]{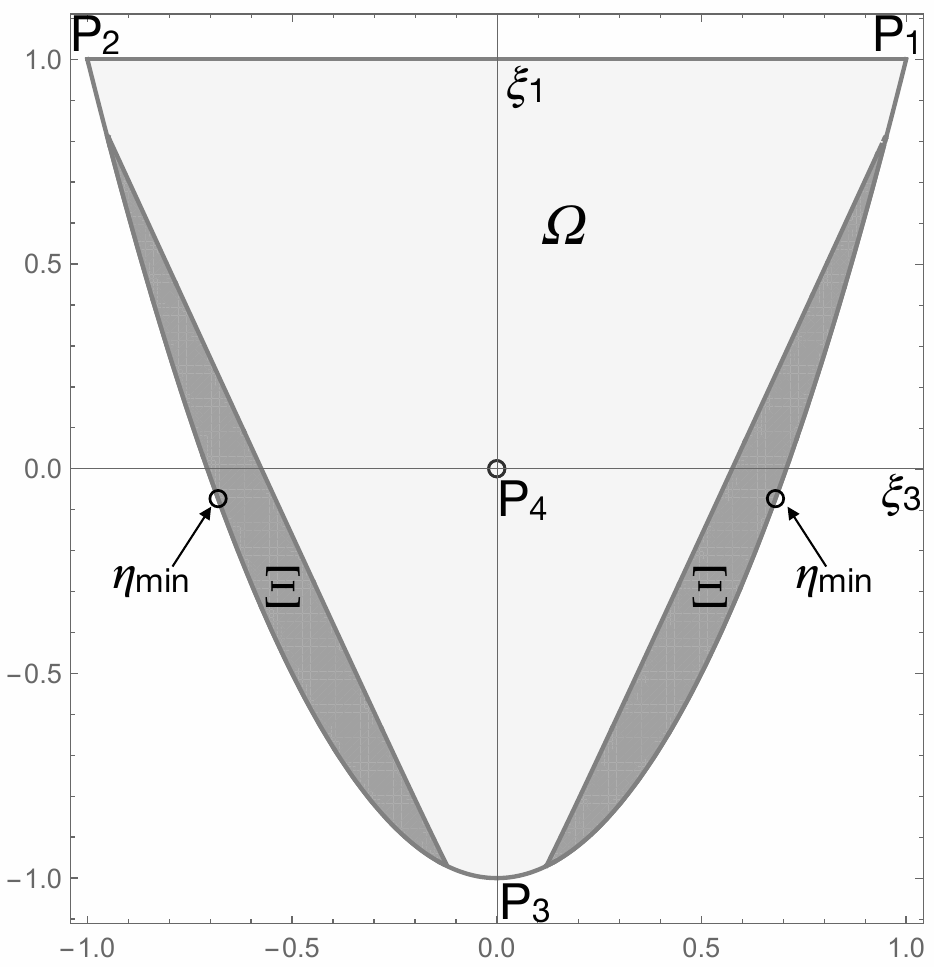}
\caption{The lamination domain $\Omega$ for an orthotropic tensor $\A$ and the auxeticity domain $\Xi$ for material 2 in Table \ref{tab:1}. The lamination points indicated by $\eta_{\min}$ give  the minimum of the function $\eta(\xi_3,\xi_1)$ in Eq. \ref{eq:eta}.}
\label{fig:1}
\end{figure}

The set of admissible lamination parameters $\Omega$ is bounded; in particular, Miki\cite{Miki82,Miki83} has shown that when the laminate is designed to be orthotropic in extension and uncoupled,    $\Omega$ is the set of points of the plane $(\xi_3,\xi_1)$ bounded by the inequalities
\be
\label{eq:boundaries}
2\xi_3^2-1\leq\xi_1\leq1,\ \ -1\leq\xi_1\leq1.
\ee
Each lamination point $(\xi_3,\xi_1)$ determines  a tensor $\A$, that, however, can be realized by more than one stacking sequence, in general. The domain $\Omega$ is represented in Fig. \ref{fig:1}. 
Lamination points P$_1=(1,1)$ and P$_2=(-1,1)$  correspond to a laminate with all the plies at $0$ or $\frac{\pi}{2}$ respectively, the point P$_3=(0,-1)$ to a balanced angle ply with orientation angles $\pm\frac{\pi}{4}$, P$_4=(0,0)$ to an isotropic laminate.  In addition, the lamination point of an angle-ply laminate belongs necessarily to the parabolic boundary of $\Omega$, while that of a  cross-ply laminate to the line P$_1-$P$_2$. 

Miki\cite{Miki85} has also shown that a tensor $\D$ describing the bending behavior has exactly the same lamination domain, although the definition of the relevant lamination parameters is different. That is why all the results of this paper, concerning extension, can be exported identically to bending, the stacking sequence apart. We just remark  the geometric meaning of a negative Poisson's ratio for $\D$: the curvatures $\kappa_1$ and $\kappa_2$ due to a bending moment $M_1$ have the same sign, i.e. locally the deformed surface is made of elliptic points, cf. \cite{toponogov,pressley,vannucci_alg}.

\subsection{Auxeticity conditions for an orthotropic laminate}
In \cite{vannucci24b} it has been shown that an orthotropic laminate can be auxetic for some directions if and only if 
\be\min_{(\xi_3,\xi_1)\in\Omega}\ \eta(\xi_3,\xi_1)<0,
\ee
with
\be
\label{eq:eta}
\eta(\xi_3,\xi_1):=2(\tau_0\tau_1-\xi_3^2)-\tau_0^2+\rho^2\xi_1^2-2|\xi_3^2\hspace{-1mm}-\hspace{-1mm}(-1)^K\tau_1\rho\ \xi_1|.
\ee
Whether or not this condition is satisfied for some lamination points, i.e. on  a subset $\Xi\subset\Omega$, depends  upon the material properties, i.e. on $\tau_0,\tau_1$ and $(-1)^K\rho$. Hence, auxetic laminates can be obtained only using plies with some specific properties. The discussion of the minimum of $\eta(\xi_3,\xi_1)$ and hence of the auxeticity condition is rather articulated and for that the reader is addressed to \cite{vannucci24b}. Here, we just recall that  the minimum of $\eta(\xi_3,\xi_1)$ belongs to the parabolic boundary of $\Omega$, i.e. it can be realized by angle-ply sequences, but not exclusively. 

Table \ref{tab:1} shows the characteristics of some UD composite plies whose   mechanical properties give a non empty subset $\Xi$ of $\Omega$. The subset $\Xi$ and the lamination points giving the minimum of $\eta(\xi_3,\xi_1)$ for material 2 in Table \ref{tab:1} are shown in Fig. \ref{fig:1}.

\begin{table}
\centering\footnotesize%\sf
\caption{Some examples of UD plies. Modules are in GPa, $\Phi_0=\Phi_1=0\Rightarrow K=0$ for all the plies.}
\begin{tabular}{rrrrrrrrrrrr}
\toprule
Mat.&$E_1$&$E_2$&$G_{12}$&$\nu_{12}$&$T_0$&$T_1$&$R_0$&$R_1$&$\tau_0$&$\tau_1$&$\rho$\\
\midrule
1&10.00&0.42&0.75&0.24&1.66&1.34&0.91&1.20&1.383&1.116&0.758\\
2&181.00&10.30&7.17&0.28&26.88&24.74&19.71&21.43&1.254&1.154&0.919\\
3& 205.00&18.50&5.59&0.23&29.80&29.14&24.21&23.42&1.272&1.244&1.033\\
4&86.90&5.52&2.14&0.34&12.23&12.11&10.09&10.25&1.193&1.181&0.984\\
5&207.00&5.00&2.60&0.25&27.52&26.85&24.93&25.29&1.088&1.062&0.986\\
6&76.00&5.50&2.10&0.34&10.85&10.74&8.75&8.88&1.221&1.209&0.984\\
7&207.00&21.00&7.00&0.30&30.67&30.35&23.67&23.46&1.307&1.293&1.009\\
8&134.00&7.00&4.20&0.25&19.34&18.12&15.14&15.93&1.214&1.138&0.951\\
9&85.00&5.60&2.10&0.34&11.98&11.89&9.88&10.00&1.198&1.189&0.988\\
10&294.50&6.34&4.90&0.23&39.73&38.01&34.83&36.06&1.102&1.054&0.966\\
11&109.70&8.55&5.31&0.30&16.89&15.53&11.58&12.73&1.327&1.220&0.910\\
12&131.70&8.76&5.03&0.28&19.55&18.26&14.52&15.45&1.265&1.182&0.940\\
13&133.10&9.31&3.74&0.34&19.02&18.74&15.28&15.60&1.219&1.201&0.979\\
14&135.00&9.24&6.28&0.32&20.55&18.89&14.27&15.83&1.298&1.193&0.901\\
15&128.00&13.00&6.40&0.30&20.00&18.77&13.60&14.51&1.378&1.293&0.937\\
\bottomrule
\end{tabular}\\
\begin{footnotesize}
\begin{tabular}{ll}
1: Pine wood\cite{Lekhnitskii}&9\hspace{1.6mm}: Kevlar-epoxy\cite{gay14}\\
2: Carbon-epoxy T300/5208\cite{TsaiHahn}&10: Carbon-epoxy GY70/34\cite{MILHDBK}\\
3: Boron-epoxy B(4)-55054\cite{TsaiHahn}&11: Carbon-bismaleimide AS4/5250-3\cite{MILHDBK}\\
4: Kevlar-epoxy 149\cite{daniel94}&12: Carbon-peek AS4/APC2\cite{MILHDBK}\\
5: Carbon-epoxy\cite{Jones}             &13: Carbon-epoxy AS4/3502\cite{MILHDBK}\\
6: Kevlar-epoxy\cite{Jones}                    &14: Carbon-epoxy T300/976\cite{MILHDBK}\\
7: Boron-epoxy\cite{Jones}                &15: Carbon-epoxy 3 MXP251S\cite{Gurdal99}\\
8: Carbon-epoxy\cite{gay14}            &\\
\end{tabular}
\end{footnotesize}
\label{tab:1}
\end{table}

\section{Minimization of $\nu_{12}(\theta)$}
\begin{figure}
\centering
\includegraphics[width=.5\columnwidth]{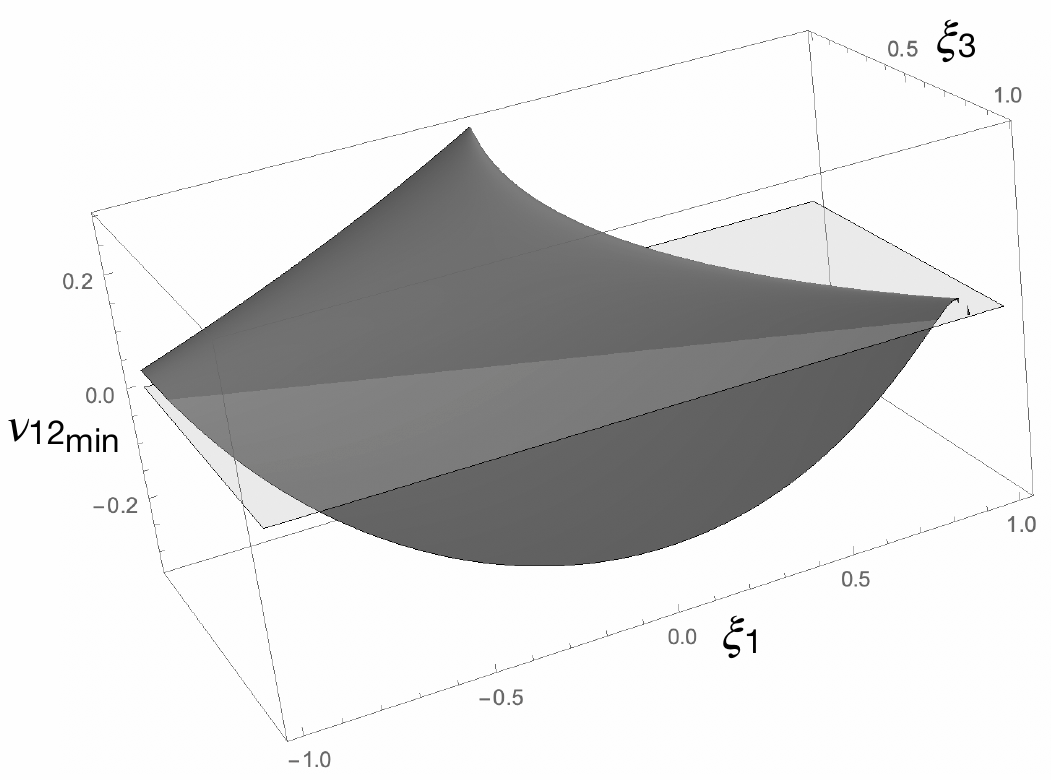}
\caption{The function ${\nu_{12}}_{\min}(\xi_3,\xi_1)$ for material 2 in Tab. \ref{tab:1}.}
\label{fig:2}
\end{figure}

The first problem considered in this paper is: for a giving layer, determine the lamination point giving the laminate having the lowest Poisson's ratio. Mathematically speaking, this amounts to solve the following constrained nonlinear optimization problem:
\be
\begin{split}
&\ \mathrm{given\  }\tau_0,\tau_1,(-1)^K\rho,\\ &\min_{(\xi_3,\xi_1)\in\Xi,\ \theta\in\left[0,\frac{\pi}{2}\right]}\nu_{12}^A(\theta),
\end{split}
\ee
with $\nu_{12}^A(\theta)$ given by Eq. (\ref{eq:nud4}). Unfortunately, this minimization problem does not have a solution in an analytical form, the equations being too involved. Nevertheless, some considerations can be done analyzing what actually happens for the function $\nu_{12}^A(\theta)$ in the lamination domain $\Omega$. To this end, setting $\frac{d\nu_{12}^A(\theta)}{d\theta}=0$,  some standard passages give the root
\be
\cos2\theta=-\frac{B+\sqrt{B^2-4C(A-C)}}{4C},
\ee
where
\be
\begin{array}{c}
A=\xi_3[\rho^2\xi_1^2-5\xi_3^2+3(-1)^K\rho\tau_1\xi_1+2\tau_0\tau_1],\bigskip\\
B=-2[(-1)^K\rho\xi_1+\tau_0][(-1)^K\rho\tau_1\xi_1-\xi_3^2],\bigskip\\
C=\xi_3[(-1)^K\rho\tau_1\xi_1-\xi_3^2].
\end{array}
\ee
Injecting this root into the Eq. (\ref{eq:nud4}) we obtain the variation of the minimum of $\nu_{12}(\theta)$ over the lamination domain $\Omega$; the final formula giving ${\nu_{12}}_{\min}(\xi_3,\xi_1)$ is omitted here because too much long. However, this function is plotted in Fig. \ref{fig:2} for  material 2 in Tab. \ref{tab:1}, where it is apparent that the lamination point giving the minimum of ${\nu_{12}}_{\min}(\xi_3,\xi_1)$ is located on the parabolic boundary of $\Omega$. 

\begin{figure}
\centering
\includegraphics[width=.5\columnwidth]{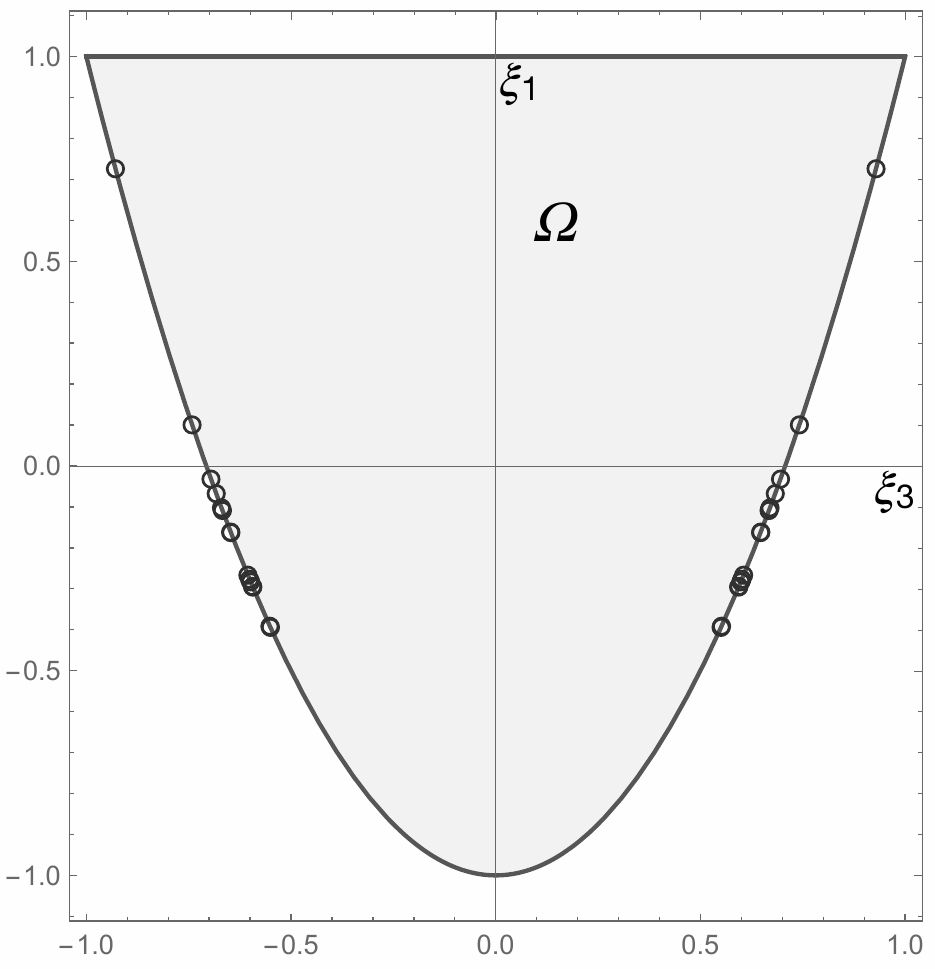}
\caption{The  lamination points giving the minimum $\nud$ for all the materials in  Table \ref{tab:1}. The points on the left side of the boundary corresponds to angle-ply laminates with the orientation angles turned of $90^\circ$ with respect to the orientations of the corresponding points on the right side.}
\label{fig:3}
\end{figure}

This situation is quite typical. In fact, an investigation made using a % numerical minimization algorithm, in this case a PSO algorithm able to handle constrained optimization problems, \cite{vannucci09algo} 
standard numerical search algorithm, confirms that for all the materials in Table \ref{tab:1} the optimal lamination point is located on the parabolic boundary of $\Omega$, see Fig. \ref{fig:3}. 

An interesting consequence of these results is that for all these materials the orthotropic  laminate with the minimum $\nud$ can be obtained using an angle-ply sequence. This generalizes  what argued by Miki and Murotsu, \cite{Miki89}: two directions, namely $\pm\delta%(\tau_0,\tau_1,(-1)^k\rho)
$, are sufficient to obtain, for a given UD material, the orthotropic laminate with the lowest (negative) Poisson's ratio. Finally, we can formulate a conjecture: {\it for any UD material, the orthotropic laminate having the most negative, i.e. with the lowest, Poisson's ratio, can always be obtained using an angle-ply sequence}. A rigorous, definitive mathematical demonstration of this conjecture remains  to be done, but apparently it seems, for the while, impossible to be get.

To end this Section, in Tab. \ref{tab:2} the minimum value of ${\nu_{12}}_{\min}(\theta)$ for all the materials in Tab. \ref{tab:1} is presented, along with the coordinates $(\xi_3,\xi_1)$ of the lamination point and the angle $\delta$ of the corresponding angle-ply laminate. To remark that, in all the cases, $\theta<45^\circ$ and $\delta<30^\circ$.

\begin{table}
\centering\footnotesize%\sf
\caption{Minimum Poisson's ratio for laminates composed of the materials in Table \ref{tab:1}.}
\begin{tabular}{rrrrrr}
\toprule
Mat.&${\nud}_{\min}$&$\theta\ (^\circ)$&$\xi_3$&$\xi_1$&$\delta\ (^\circ)$\\
\midrule
1&-0.42&31.2&0.93&0.72&10.9\\
2&-0.33&39.1&0.68&-0.07&23.5\\
3&-0.23&41.9&0.55&-0.39&28.3\\
4&-0.35&40.2&0.60&-0.27&26.4\\
5&-0.95&34.2&0.69&-0.03&23.0\\
6&-0.28&40.9&0.59&-0.29&26.8\\
7&-0.16&42.7&0.55&-0.39&28.3\\
8&-0.39&38.6&0.67&-0.11&24.1\\
9&-0.33&40.4&0.60&-0.28&26.6\\
10&-0.94&33.1&0.74&0.10&21.0\\
11&-0.20&41.2&0.65&-0.16&24.8\\
12&-0.28&40.2&0.65&-0.16&24.8\\
13&-0.29&40.7&0.60&-0.27&26.5\\
14&-0.24&40.3&0.67&-0.10&24.0\\
15&-0.12&42.8&0.59&-0.29&26.8\\
\bottomrule
\end{tabular}\\
\label{tab:2}
\end{table}

\section{Maximization of the auxetic zone}
In a previous work it has been proved that a totally auxetic orthotropic laminate composed by orthotropic plies that are not auxetic themselves cannot exist, \cite{vannucci24b}. As a consequence, it is meaningful and interesting for practical applications to find the conditions that maximize the {\it auxetic zone} of such a laminate. In other words, for a given material, we look for the stacking sequence giving the laminate that has the largest set of directions where $\nud<0$.

\begin{figure}
\centering
\includegraphics[width=.6\columnwidth]{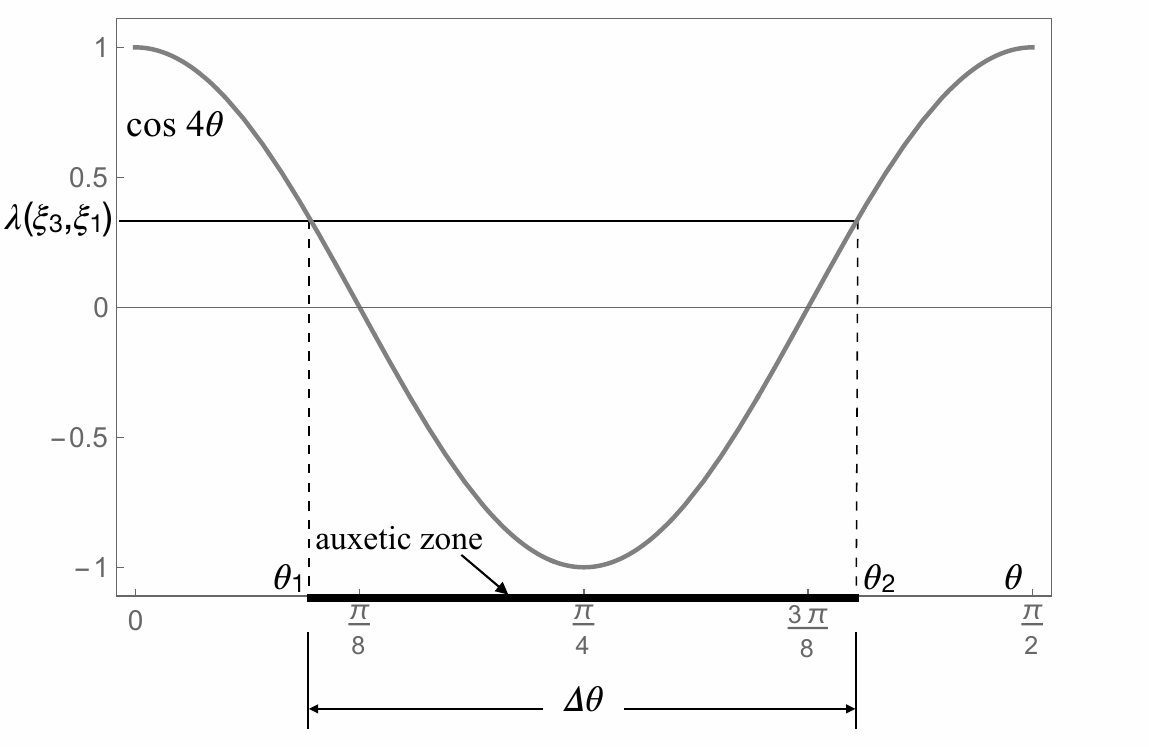}
\caption{Symmetry of the  auxetic zone.}
\label{fig:4}
\end{figure}

Because $\Sy_{11}(\theta)>0\ \forall\theta,$ the sign of $\nud(\theta)$ depends uniquely on the numerator in Eq. (\ref{eq:nud}) and hence, Eq.  (\ref{eq:nud4}), $\nud^A(\theta)>0\iff$
\be
\psi(\theta):=2(\tau_0\tau_1{-}\xi_3^2){-}\tau_0^2{+}\rho^2\xi_1^2{+}2[\xi_3^2{-}(-1)^K\tau_1\rho \xi_1]\cos4\theta<0.
\ee
The condition of auxeticity for an orthotropic laminate is hence periodic of $\pi/2$, which means that if a material allows the fabrication of partially auxetic orthotropic laminates, \cite{vannucci24b}, then there are four auxetic zones symmetric with respect to the $x_1$ and $x_2$ axes and that $\nud>0$ in correspondence of the axes, i.e. for $\theta=0,\pi/2$. In fact, 
\be
\label{eq:psi1}
\psi(\theta)<0\ \iff\ \cos4\theta<\lambda(\xi_3,\xi_1):=\frac{\tau_0^2-\rho^2\xi_1^2-2(\tau_0\tau_1{-}\xi_3^2)}{2[\xi_3^2-(-1)^K\tau_1\rho \xi_1]},
\ee
and because $\cos4\theta$ is periodic of $\pi/2$, the auxetic zone, from now on denoted by $\Delta\theta$, is necessarily symmetric with respect to  $\theta=\pi/4$, see Fig. \ref{fig:4}. 
For $\theta=0,\pi/2$  it is necessarily $\nud>0$ because, as shown in \cite{vannucci24b}, totally auxetic orthotropic laminates made by identical non-auxetic orthotropic layers cannot exist.

\begin{figure}
\centering
\includegraphics[width=.5\columnwidth]{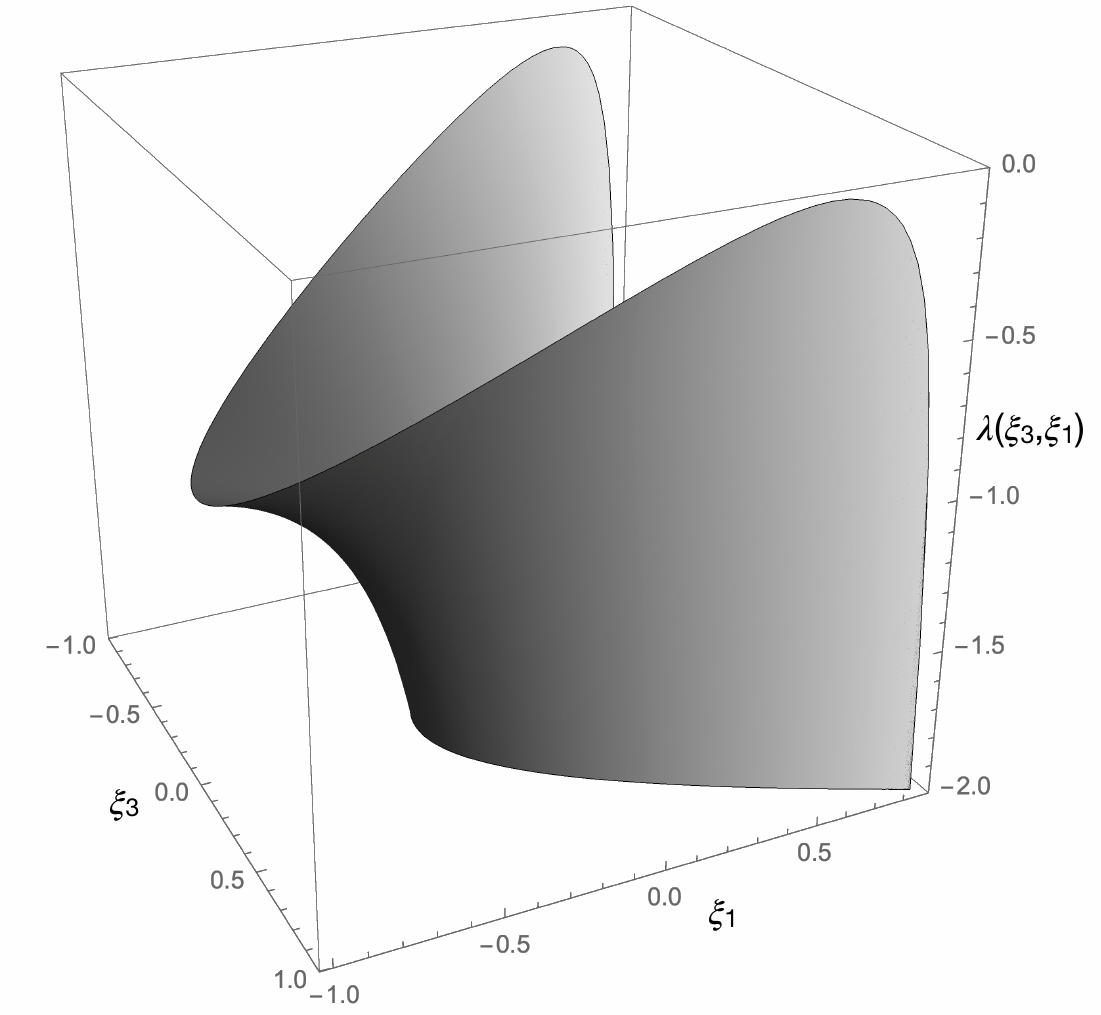}
\caption{The function $\lambda(\xi_3,\xi_1)$ for material 2 in Tab. \ref{tab:1}.}
\label{fig:5}
\end{figure}

The problem considered in this Section, i.e. the maximization of the interval $\Delta\theta$,  corresponds hence to:
\be
\max_{(\xi_3,\xi_1)\in\Xi}\lambda(\xi_3,\xi_1).
\ee
Unlike the case studied in the previous Section, this problem has an analytical solution. In fact, differentiating $\lambda(\xi_3,\xi_1)$ with respect to $\xi_3$ gives
\be
\frac{\partial\lambda}{\partial\xi_3}=\xi_3\frac{\rho^2\xi_1^2-2(-1)^K\tau_1\rho\xi_1-\tau_0(\tau_0-2\tau_1)}{4\left[\xi_3^2-(-1)^K\tau_1\rho\xi_1\right]^2}.
\ee
It is hence apparent that $\frac{ \partial\lambda(0,\xi_1)}{\partial\xi_3}=0$, i.e. for a given $\xi_1=\xi_1^*$, the function $\lambda(\xi_3,\xi_1^*)$ has a stationary point only at $\xi_3=0$, then it is necessarily positive or negative $\forall(\xi_3,\xi_1^*)\in\Xi$.  %then, because $\lambda(\xi_3,\xi_1)$ is an even function of $\xi_3$, cf. Eq. (\ref{eq:psi1}), this stationary point can be only a maximum or a minimum for $\lambda(\xi_3,\xi_1^*)$. Actually, it cannot be a maximum,  
Then, because the line $\xi_3=0\notin\Xi$, otherwise partially auxetic laminates could exist $\forall(\xi_3,\xi_1)\in\Omega$, which cannot be, see \cite{vannucci24b}, along any straight line $\xi_1=\xi_1^*$ the maximum of $\lambda(\xi_3,\xi_1^*)$ is get for the maximum value of $\xi_3$, i.e. on the parabolic boundary of $\Omega$, cf. Fig. \ref{fig:1} (on the boundary of $\Xi, \nud=0\Rightarrow\Delta\theta=0$). A typical example of the function $\lambda(\xi_3,\xi_1)$ is shown in Fig. \ref{fig:5} for material 2 in Tab. \ref{tab:1}.

\begin{figure}
\centering
\includegraphics[width=.5\columnwidth]{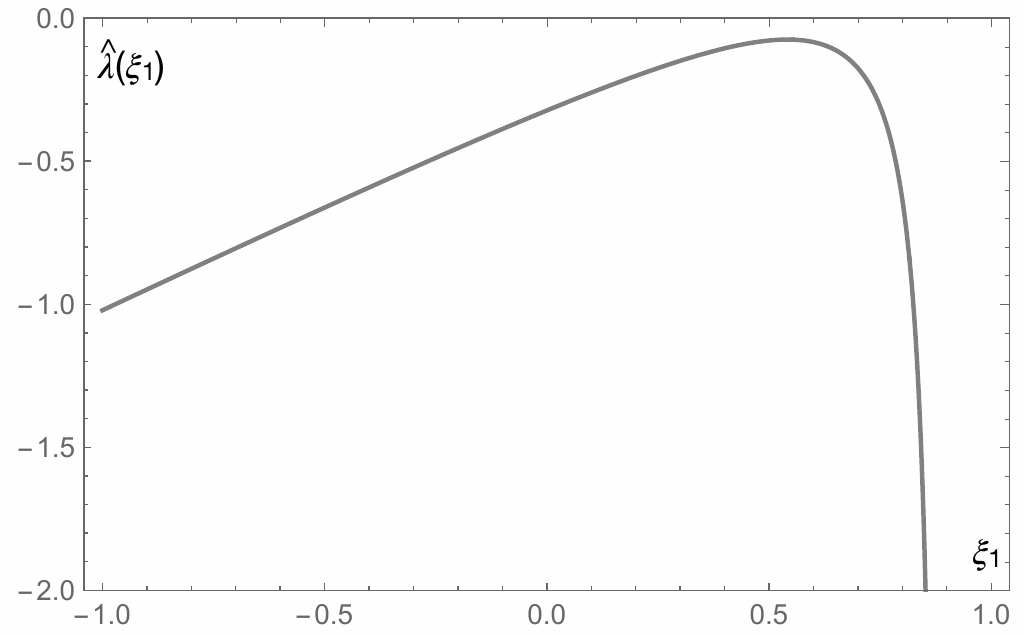}
\caption{The function $\widehat{\lambda}(\xi_1)$ for material 2 in Tab. \ref{tab:1}.}
\label{fig:6}
\end{figure}

On such parabolic boundary, see Eq. (\ref{eq:boundaries}), 
\be
\xi_3^2=\frac{1}{2}(\xi_1+1),
\ee
we put $\lambda(\xi_3,\xi_1)=\widehat{\lambda}(\xi_1)$ with
\be
\widehat{\lambda}(\xi_1)=\frac{\tau_0^2-2\tau_0\tau_1+1+\xi_1-\rho^2\xi_1^2}{1+[1-2(-1)^K\tau_1\rho] \xi_1}.
\ee
The function $\widehat{\lambda}(\xi_1)$ for material 2 in Tab. \ref{tab:1} is shown in Fig. \ref{fig:6}.  
Differentiating, after some standard passages, we get that $\frac{d\widehat{\lambda}}{d\xi_1}=0$ for
\be
\xi_1^{opt}=\frac{\rho^2-\sqrt{\rho^4+\rho^2(2(-1)^K\rho\tau_1-1)[(2(-1)^K\rho\tau_1-1)(2\tau_1-\tau_0)\tau_0-2(-1)^K\rho\tau_1]}}{\rho^2(2(-1)^K\rho\tau_1-1)},
\ee
which gives also 
\be
\xi_3^{opt}=\pm\sqrt{\frac{\xi_1^{opt}+1}{2}}.
\ee
The lamination point $(\xi_3^{opt},\xi_1^{opt})$ maximizes $\Delta\theta$ for the given material. Because this point is on the parabolic boundary of $\Omega$, it can be realized fabricating an angle-ply laminate having orientation angles $\pm\delta$ with
\be
\delta=\frac{1}{2}\arccos\xi_3^{opt}=\frac{1}{4}\arccos\xi_1^{opt}.
\ee

\begin{figure}
\centering
\includegraphics[width=.5\columnwidth]{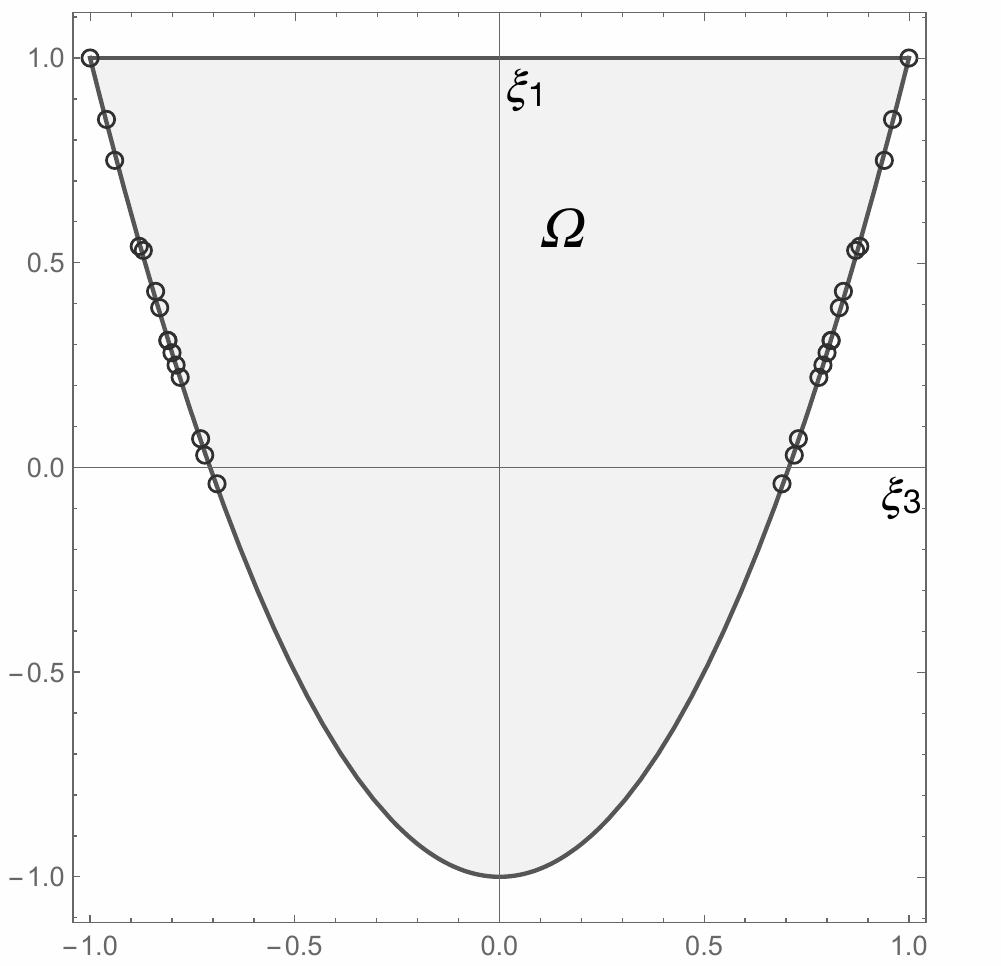}
\caption{The  lamination points giving the maximum of $\Delta\theta$ for all the materials in  Table \ref{tab:1}. The points on the left side of the boundary corresponds to angle-ply laminates with the orientation angles turned of $90^\circ$ with respect to the orientations of the corresponding points on the right side.}
\label{fig:7}
\end{figure}

The situation depicted in Figures \ref{fig:5} and \ref{fig:6} is quite typical and  
 in Tab. \ref{tab:3} the  value of $\Delta\theta$ is shown for all the materials in Tab. \ref{tab:1}; the table is completed  with the coordinates $(\xi_3^{opt},\xi_1^{opt})$ of the lamination point, the angles $\theta_1$ and $\theta_2$, bounds of the auxetic zone, see Fig. \ref{fig:4}, the angle $\delta$ of the corresponding angle-ply laminate, the value of ${\nu_{12}}_{\min}(\theta)$ and its direction $\theta_{\min}$. 
To remark that, unlike the previous case, $\theta_{\min}$ can be greater than $45^\circ$, while  in all the cases $\delta<30^\circ$, like before. Moreover, in the most part of cases it is  $\max\lambda(\xi_3,\xi_1)<0\Rightarrow\Delta\theta<\frac{\pi}{4}$, see Fig. \ref{fig:4}. However, this cannot be generalized: for materials 1, 5 and 10, $\max\lambda(\xi_3,\xi_1)>0$ and $\Delta\theta>\frac{\pi}{4}$. The lamination points giving $\Delta\theta_{\max}$ for all the materials in Tab. \ref{tab:1} are shown in Fig. \ref{fig:7}. Comparing this figure with Fig. \ref{fig:3}, it is apparent that the lamination point giving $\Delta\theta_{\max}$ are, generally speaking, located in a position on the boundary of $\Omega$ having a greater $\xi_1$ than the points giving, for the same material, the minimum of $\nud$. In practice, this means that the angle-ply giving $\Delta\theta_{\max}$ has an orientation angle $\delta$ smaller than the one giving the minimum of $\nud$. Comparing the results in Tables \ref{tab:2} and \ref{tab:3}, it can be checked that this is true for all the examined materials.

\begin{table}
\centering\footnotesize%\sf
\caption{Maximum auxetic zone for laminates composed of the materials in Table \ref{tab:1}.}
\begin{tabular}{rrrrrrrrr}
\toprule
Mat.&$\xi_3^{opt}$&$\xi_1^{opt}$&$\theta_1\ (^\circ)$&$\theta_2\ (^\circ)$&$\Delta\theta\ (^\circ)$&$\delta\ (^\circ)$&${\nud}_{\min}$&$\theta_{\min}\ (^\circ)$\\
\midrule
1&1.00&1.00&8.9&81.1&72.2&0.0&-0.39&27.2\\
2&0.88&0.54&23.6&66.4&42.8&14.3&-0.19&37.2\\
3&0.73&0.07&30.7&59.3&28.6&21.5&-0.16&41.1\\
4&0.81&0.31&26.9&63.1&36.2&18.0&-0.21&38.9\\
5&0.94&0.75&16.2&73.8&57.6&10.3&-0.36&30.5\\
6&0.78&0.22&28.6&61.4&32.7&19.3&-0.18&39.9\\
7&0.69&-0.04&33.0&56.9&23.9&23.1&-0.12&42.3\\
8&0.87&0.53&22.9&67.0&44.0&14.5&-0.21&36.6\\
9&0.80&0.28&27.4&62.6&35.1&18.5&-0.21&39.2\\
10&0.96&0.85&13.9&76.0&62.1&8.0&-0.31&28.9\\
11&0.81&0.31&28.3&61.7&33.3&17.9&-0.14&40.2\\
12&0.83&0.39&26.2&63.8&37.6&16.8&-0.17&38.8\\
13&0.79&0.25&28.1&61.9&33.8&18.8&-0.19&39.7\\
14&0.84&0.43&26.3&63.7&37.3&16.2&-0.16&39.0\\
15&0.72&0.03&32.9&57.1&24.1&22.1&-0.10&42.4\\
\bottomrule
\end{tabular}\\
\label{tab:3}
\end{table}

\section{Numerical examples and final considerations}

\begin{figure}
\centering
\includegraphics[width=.7\columnwidth]{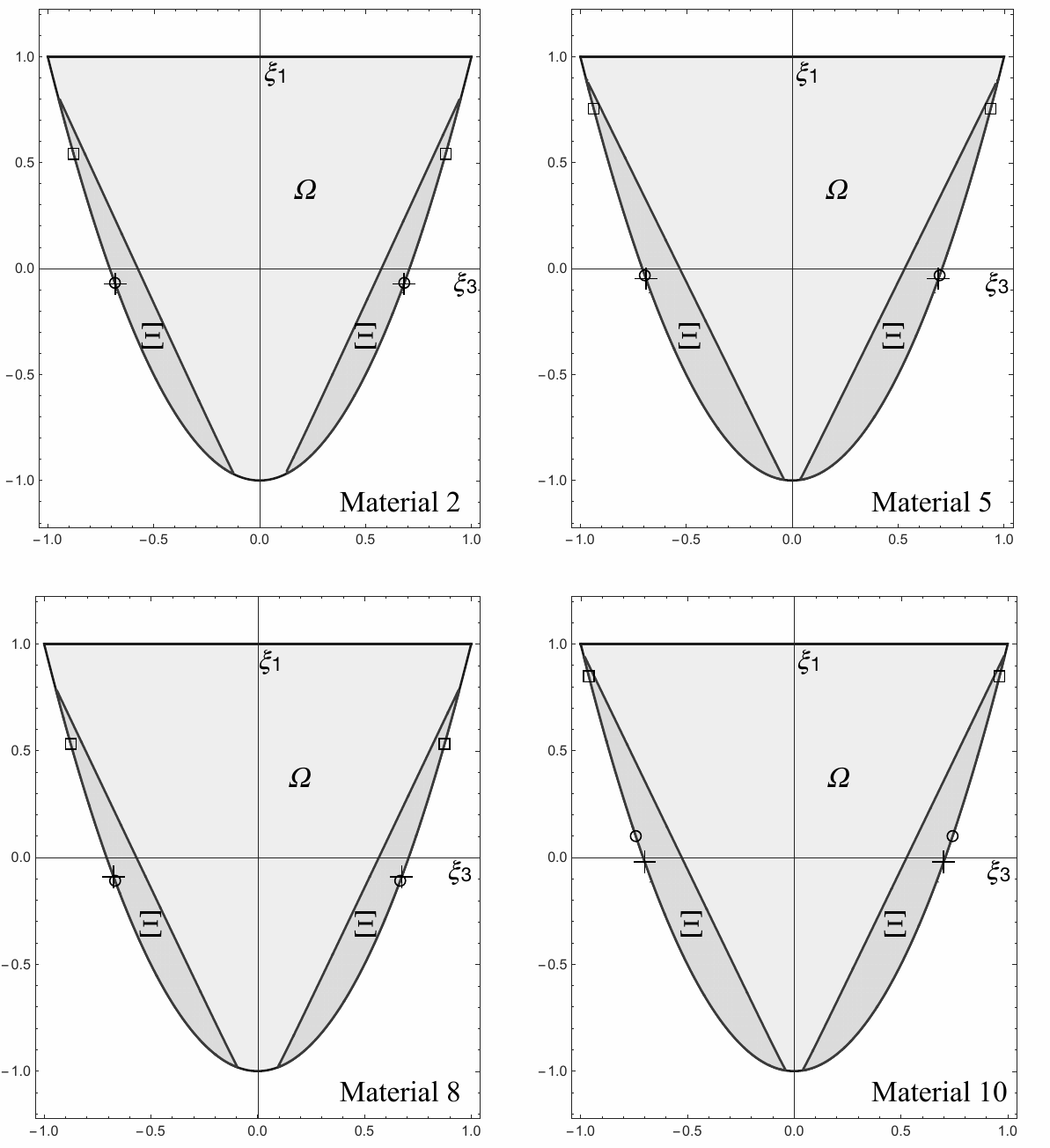}
\caption{The case of four materials in Tab. \ref{tab:1}: for each material, the domain $\Xi\subset\Omega$ giving partially auxetic laminates is indicated, along with the lamination points corresponding to the laminates with $\eta_{\min},{\nud}_{\min}, \Delta\theta_{\max}$, denoted respectively by $\circ,+$ and {\footnotesize$\square$}.}
\label{fig:8}
\end{figure}

\begin{figure}
\centering
\includegraphics[width=.7\columnwidth]{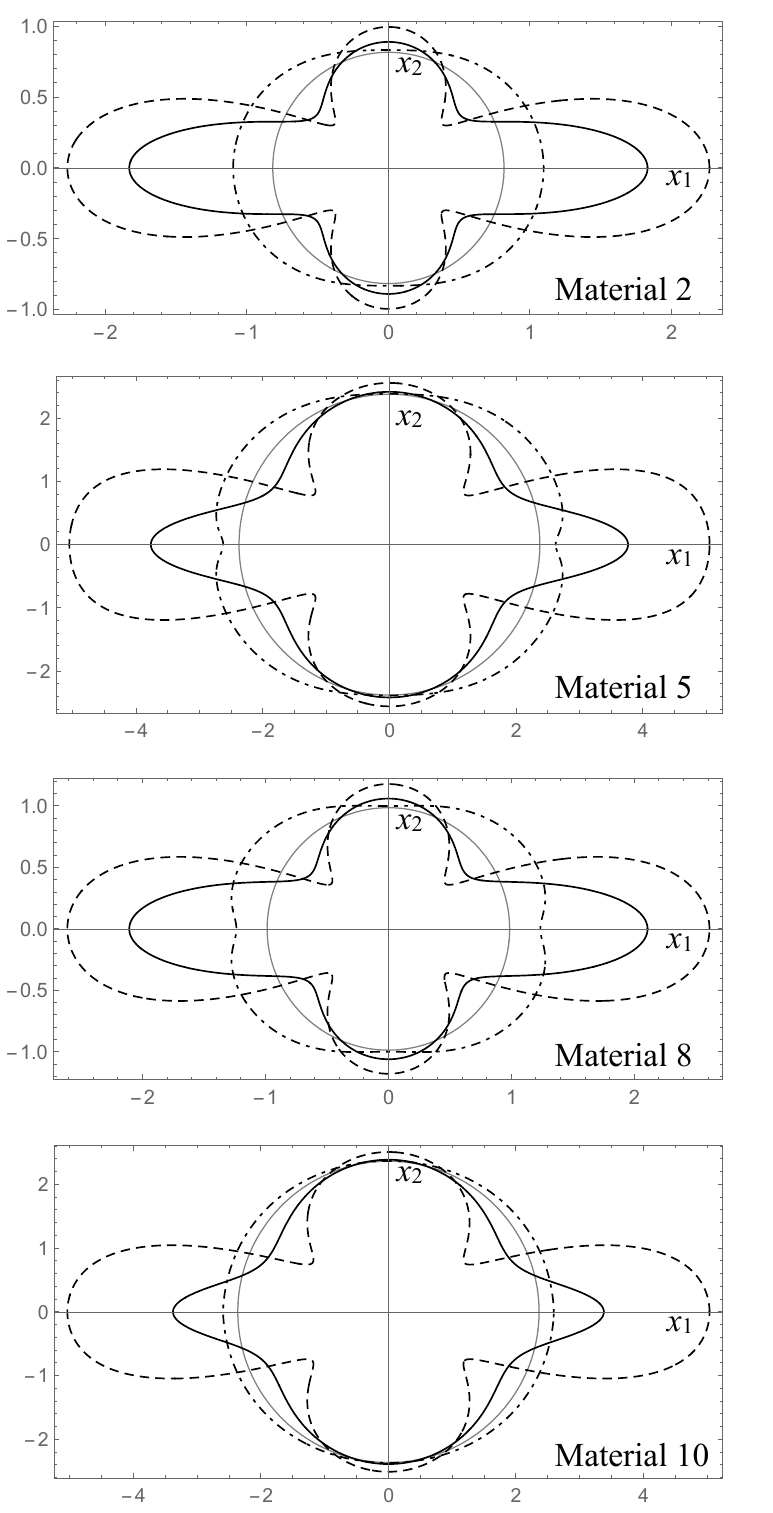}
\caption{The directional diagrams of $\nud$ for four materials in Tab. \ref{tab:1}: for each material, the curves representing $\nud$ of the single layer and of the laminates  with ${\nud}_{\min}$ and $ \Delta\theta_{\max}$ are shown, denoted respectively by a dot-dashed, a dashed and a solid line. The thin circle marks the zero value: inside it, $\nud$ is negative.}
\label{fig:9}
\end{figure}

For example, we consider in this Section four materials among those in Tab. \ref{tab:1} and present for them the maps of the lamination points, Fig. \ref{fig:8},  and the directional diagrams of $\nud$, Fig. \ref{fig:9},  for the laminates solution of the two problems considered above.

These four cases are quite typical of the auxeticity of laminates composed by standard composite materials. In all of these cases we can remark that the laminate with $\Delta\theta_{\max}$ is always get using an angle ply with a smaller $\delta$ than the laminate having ${\nud}_{\min}$. The difference in the auxetic zone between these two types of laminates can be rather important. Even more different is the minimum value of the Poisson's ratio: passing from the solution for  $\Delta\theta_{\max}$ to that for ${\nud}_{\min}$, the  value of $\nud$  passes from $-0.19$ to $-0.33$ for material 2, from $-0.36$ to $-0.95$ for material 5, from $-0.21$ to $-0.39$ for material 8 and from $-0.31$ to $-0.94$ for material 10. Theses data show perfectly that the laminates solution of the two different problems have, for the same constituent material, very different auxetic properties. 

To conclude, we recall that angle-ply stacks can always be used to obtain auxetic laminates, although this is not compulsory, and  remark that the auxetic properties of the laminates are rather sensitive to the orientation angle $\delta$ of the angle-ply sequence: a small change of $\delta$ can change substantially the auxeticity of the laminate.

%\section{Conclusion}

%Unlike the first problem, the second one has an analytical solution and it can be proved mathematically that the optimal solution is on the parabolic boundary of $\Omega$, while this can just remain a conjecture for the first problem.

\bibliographystyle{SageV}
\bibliography{Biblio.bib}

\end{document}